\documentclass[amsmath,amssymb,prc,final,showpacs,twocolumn]{revtex4}

\usepackage{bm,hyphenat,xspace} \usepackage{graphicx,epsfig}

\newcommand{\scl}{0.63}
  
\newcommand{\Eq}{Eq.}  \newcommand{\Eqs}{Eqs.}
\newcommand{\Fig}{Fig.}  \newcommand{\Figs}{Figs.}
\newcommand{\Ref}{Ref.}  \newcommand{\Refs}{Refs.}

\newcommand{\cm}{\mathrm{c\!\:\!.m\!\:\!.}}
\newcommand{\C}{{}^{12}\mathrm{C}}
\newcommand{\Cn}{{}^{13}\mathrm{C}}
\newcommand{\A}[2]{{}^{#1}\mathrm{#2}}
\newcommand{\Ox}{{}^{16}\mathrm{O}}
\newcommand{\On}{{}^{17}\mathrm{O}}

\newcommand{\np}{np} 
\newcommand{\Bee}{{}^{11}\mathrm{Be}}
\newcommand{\etal}{{\em et al.}}

\begin{document}
 
\title {Three-body Faddeev-Alt-Grassberger-Sandhas approach to direct nuclear reactions}
 
\author{A.~Deltuva} 
\affiliation{Centro de F\'{\i}sica Nuclear da Universidade de Lisboa,
  P-1649-003 Lisboa, Portugal }

\author{A.~C.~Fonseca}  \affiliation{Centro de F\'{\i}sica Nuclear da
  Universidade de Lisboa,  P-1649-003 Lisboa, Portugal }

\received{October 30, 2008}

\begin{abstract}
Momentum space three-body Faddeev-like equations are used to calculate 
elastic, transfer and charge exchange reactions resulting from the 
scattering of deuterons on $\C$ and $\Ox$ or protons on $\Cn$ and $\On$; 
$\C$ and $\Ox$ are treated as inert cores. 
All possible reactions are calculated in the framework of the same model 
space. Comparison with previous calculations based on approximate methods 
used in nuclear reaction theory is discussed.
\end{abstract}

 \maketitle
 

\section{Introduction \label{sec:intro}}

As discussed in the review article by
Austern~{\etal}~\cite{Austern-87} twenty years ago, three-body models
of deuteron-induced reactions became important since the early studies
of stripping theory~\cite{Butler51}, where ``the internal coordinates
of the target nucleus are ignored and the only dynamically active
variables are the coordinates, relative to the target nucleus, of the
interacting nucleon that is captured by the nucleus and the spectator
nucleon that goes on to the detector".

The present work goes back in time, recaptures the three-body concept
of direct nuclear reactions that is common to continuum discretized
coupled channels (CDCC) calculations~\cite{Austern-87}  and shows the
results obtained by solving Faddeev/Alt, Grassberger, and Sandhas
(AGS) equations~\cite{Faddeev60,Alt-67,Glockle83} for elastic,
transfer and breakup reactions where three-body dynamics plays a dominant 
role. In this work we attempt to calculate all observables using dynamical 
models based on energy-independent or energy-dependent optical potentials 
for the nucleon-nucleus interaction~\cite{Watson69} and realistic 
neutron-proton $(\np)$ potentials such as CD-Bonn~\cite{CDBONN}. 
Some examples are shown for reactions initiated
by deuterons on $\C$ and $\Ox$, as well as protons on $\Cn$ and $\On$.
Although the use of energy-dependent potentials in three-body
calculations is not free of theoretical problems that are
discussed below, the results we show demonstrate the possibilities
and the shortcomings of this model;
this is, above all, the aim of the present paper. In addition we present the
exact derivation of an alternative set of equations that may serve as
the basis for future investigations on improving approximate methods
in nuclear reaction theory.

Although deuteron-nucleus three-body models, including
striping or pick up, have already been explored in the past
in the framework of Faddeev/AGS equations
starting with the pioneer work of Aaron and Shanley~\cite{Aaron66} to
the more recent calculations of Alt \etal~\cite{Alt07}, all of them
were drastically simplified. In most cases separable interactions were
used between pairs and the correct treatment of the Coulomb
interaction was missing. This situation has now changed due
to the recent progress in the description of proton-deuteron elastic 
scattering and breakup~\cite{deltuva:05a,deltuva:05c}
where the Coulomb repulsion is fully included using the method of 
screening and renormalization \cite{taylor:74a,alt:78a}
together with realistic nuclear potentials.
This technical development was applied
to three-body nuclear reactions to test the accuracy of the CDCC
method~\cite{deltuva:07d} and the convergence of the multiple scattering 
series in the framework of the Glauber approximation~\cite{Crespo07b}
and distorted-wave impulse approximation (DWIA)~\cite{Crespo2008}  which 
are standard approximations used to describe nuclear reaction data. 

Some of the interaction models employed in this work and in CDCC calculations
 are formally  identical, but instead  of solving the three-body
Schr\"{o}dinger equation in coordinate space using a representation in
terms of a  set of eigenstates pertaining to a given
subsystem Hamiltonian, we solve the Faddeev/AGS equations in momentum space
and obtain numerically well converged solution of the three-body
problem  for all reactions allowed by the chosen interactions. In
\Ref~\cite{deltuva:07d} we benchmarked the two methods and concluded that
CDCC is indeed a reliable method to calculate deuteron-nucleus elastic
and  breakup cross sections, but may not provide a sufficiently accurate
solution of the three-body problem for
transfer and breakup in one-neutron halo nucleus scattering 
from a proton target  such as $\Bee + p$ reactions.
In those cases the comparison of CDCC results with experimental data
may be misleading.

In Sec.~\ref{sec:3Beq} we recall the Faddeev/AGS equations, 
in Sec.~\ref{sec:TDM} we present the results for three dynamical models,
and in Sec.~\ref{sec:DWE} we compare them with the results of standard 
approximations used in nuclear reaction theory. 
Conclusions are given in Sec.~\ref{sec:CON}.


\section{The Three-Body Equations \label{sec:3Beq}}

Let's consider a system of three particles  $(\alpha = 1,2,3)$ with
kinetic energy operator $H_0$,
interacting by means of two-body potentials $v_{\alpha}$
($v_1 = v_{23}$ in the standard odd-man-out notation). The full resolvent 
\begin{gather}\label{eq:G}
G(Z) = (Z - H_0 - \sum_{\sigma} v_{\sigma})^{-1}
\end{gather}
and the channel resolvent
\begin{gather}
G_{\alpha}(Z) = (Z - H_0 - v_{\alpha})^{-1}
\end{gather}
may be related through the AGS transition operator $U_{\beta \alpha}(Z)$ as
\begin{gather}\label{eq:G(Z)}
G(Z) = \delta_{\alpha \beta} \; G_{\alpha}(Z)  + G_{\beta}(Z) U_{\beta \alpha}(Z)  
G_{\alpha}(Z).
\end{gather}
The transition operator $U_{\beta \alpha}(Z)$ satisfies the AGS equation
\cite{Alt-67}
\begin{gather}  \label{eq:Uba}
U_{\beta \alpha}(Z)  = \bar{\delta}_{\beta\alpha} \, G^{-1}_{0}(Z)  +
\sum_{\sigma}   \bar{\delta}_{\beta \sigma} \, T_{\sigma} (Z)
\, G_{0}(Z) U_{\sigma \alpha}(Z),
\end{gather}
where the summation on $\sigma$ runs from one to
three, $\bar{\delta}_{\beta\alpha} = 1- \delta_{\beta \alpha}$,
$G_0(Z) = (Z - H_0)^{-1}$ is the free resolvent, and
$T_{\alpha}(Z)$  is the two-body transition matrix (t-matrix) that obeys the
Lippmann-Schwinger equation for pair $\alpha$,
\begin{gather}\label{eq:t_alpha}
T_{\alpha}(Z)  =  v_{\alpha}  + v_{\alpha} \, G_{0}(Z) \, T_{\alpha}(Z).
\end{gather}
At a given energy $E$ in the three-body center of mass $(\cm)$ system
the on-shell matrix elements 
$\langle\psi_{\beta}|U_{\beta \alpha}(E+i0)|\psi_{\alpha}\rangle$ 
calculated between the appropriate channel states 
yield all the relevant elastic, inelastic and transfer  $(\beta, \alpha =
1, 2, 3)$ as well as breakup $(\beta = 0)$ amplitudes.
The channel state $|\psi_{\alpha}\rangle$ for  $\alpha = 1,2,3$ is the 
eigenstate of the corresponding channel Hamiltonian $H_\alpha = H_0 + v_\alpha$
with the energy eigenvalue $E$ made up by the bound state wave
function for pair $\alpha$ times a relative plane wave between
particle  $\alpha$ and pair $\alpha$. For breakup the final state is
a product of two plane waves corresponding to the relative
motion of three free particles.

The AGS equations \eqref{eq:Uba} are Faddeev-like equations
with compact kernel and therefore suitable for numerical solution;
they are consistent with the corresponding Schr\"{o}dinger equation
and therefore provide exact description of the quantum three-body problem.
After partial wave decomposition \Eq~\eqref{eq:Uba} becomes
a two variable integral equation which we solve by standard
discretization of momentum variables and summation of the multiple
scattering  series by Pad\'{e} method; more details can be found in
\Refs~\cite{chmielewski:03a,deltuva:03a}. As in all numerical
calculations convergence of results has to be tested vis-\`{a}-vis
number of included partial waves, mesh points and Pad\'{e} steps.

To include the Coulomb interaction between two charged particles
we use the method of screening and renormalization 
\cite{taylor:74a,alt:78a,deltuva:05a}.
The Coulomb potential is screened, 
standard scattering theory for short-range potentials is
used in the form  of \Eq~\eqref{eq:Uba} with parametric
dependence on the screening radius $R$, and the renormalization procedure 
is applied to obtain $R$-independent results for sufficiently large $R$,
that correspond to the unscreened limit. 
A complete review on this subject is presented
in \Ref~\cite{deltuva:08c} together with a number of practical applications.


\section{The Dynamical Models} \label{sec:TDM}

In this section we set the three-body dynamics we apply to study all
the reactions initiated by deuterons on $\C$ and $^{16}{\rm O}$ as
well as protons on $^{13}{\rm C}$ and $^{17}{\rm O}$ where $\C$ and
$^{16}{\rm O}$ are considered as inert cores. 

Although in most nuclear reaction calculations the deuteron wave
function is generated through a Gaussian potential fitted to the
deuteron binding energy, which is then used to drive the
$\np$ interaction in all other partial waves, we use
the CD-Bonn~\cite{CDBONN} potential as our realistic interaction for
all $\np$ partial waves including the deuteron channel. 

For the neutron-nucleus  $(nA)$ and proton-nucleus $(pA)$ interactions
we use the optical potentials of Watson~\etal~\cite{Watson69} which
are based on an optical model analysis of nucleon scattering from
$1p$-shell nuclei between 10 and 50 MeV; the  nucleus $A$ is a
structureless core of mass number $A$. Although core excitation may be
treated in the present three-body models, we discard such possibility
at this time.  Therefore the relevant parameters of this optical model
fit are both energy and mass dependent and are fitted to the existing
data over the energy and mass range. For specific nuclei and energy,
one could perhaps obtain a better fit but, as mentioned in the
Introduction, our goal is to explore the possibilities of a three-body
model that can simultaneously describe all reactions  allowed
by the chosen interactions and leave the fine tuning
for an improved model study. In all calculations nucleons are considered as
spin $1/2$ particles and the nuclear cores as spin zero particles;
the spin-orbit terms of the optical potentials are included as well
as the full operator structure of the CD-Bonn potential for the $np$
pair. The calculations include $np$ partial waves with total angular momentum
 $I \leq 3$, $nA$ partial waves with orbital angular momentum  $L \leq 8$, 
and $pA$ partial waves with   $L \leq 20$; the total three-particle
angular momentum is $J \leq 35$. Depending on the reaction and energy,
some of these quantum numbers cutoffs can be safely chosen significantly lower,
leading, nevertheless, to well converged results.
The $pA$ channel is more demanding than the $nA$ channel due to the screened 
Coulomb force, where the screening radius $R \approx 10$ fm for the
short-range part of the scattering amplitude is sufficient for 
convergence. The only exception are reactions leading to a final $(Ap)$ 
bound state where $R \approx 15$ fm and a sharper screening is needed.
With the above choice of the calculational parameters we obtain
well converged results for all considered observables such that 
all discrepancies with the experimental data can be attributed solely
to the shortcomings of the interaction models that are used.


\subsection{Model 1 - Energy-independent optical potentials}

In this case we use the traditional approach based on energy-independent 
optical potentials whose parameters are chosen at a fixed
energy. For deuteron scattering from nucleus $A$ the parameters for the 
$nA$ and $pA$ potentials are taken  from Ref.~\cite{Watson69}
at half the lab energy of the deuteron projectile. 
For proton scattering from the $(An)$ nucleus
the $pA$ parameters are taken  from Ref.~\cite{Watson69}
at the lab energy of the proton beam and
the $nA$ parameters at zero energy, where the imaginary part of the $nA$
optical potential is zero.  
Small adjustments to these $nA$ parameters are made in order to be
able to reproduce the experimental binding energies of the 
 ground and excited single particle states of the $(An)$ nucleus
while all Pauli forbidden bound states of the resulting potential are removed
as described in  \Ref~\cite{deltuva:06b}. Original \cite{Watson69}
and adjusted values of these parameters are given in Table~\ref{tab:V}.
In the present model only the modified  $nA$ parameters are used in 
given partial waves leading to the single particle states listed in
Table~\ref{tab:EB} for $\Cn$ and $\On$; 
in all other $nA$ partial waves we used the 
original  parameters \cite{Watson69} as well as for the  $pA$ optical
potential. While in $d+A$ scattering the $pA$ and $nA$
potentials are complex, in $p + (An)$ scattering only the $pA$
potential is complex. Although in both cases we are dealing with the same
particles, the Hamiltonians are different and, therefore, in  $d+A$
we cannot calculate $d+A \to p + (An)$, but  in $p+(An)$ we can calculate
the inverse reaction $p + (An) \to d + A$, or even $p + (An) \to p +
(An)^*$, because the  $nA$ interaction is real, in contrast to  $d+A$
where it is complex.

\begin{table} [htbp]
\begin{ruledtabular}
\begin{tabular}{l*{4}{c}}
 & $v_R(nA)$ & $v_R(pA)$  & $V_{so}(nA) $  & $V_{so}(pA) $ \\ \hline
 Ref.~\cite{Watson69} & 60.00 & 60.00 & 5.5  & 5.5  \\
$N$-$\C$ (s)  & 67.50 & 66.47 &   &   \\
$N$-$\C$ (p)  & 61.67 & 61.50 & 20.38  & 20.83    \\
$N$-$\C$ (d)  & 66.42 & 66.42 & 5.5  & 5.5    \\
$N$-$\Ox$ (s) & 61.65 & 60.94 &   &    \\
$N$-$\Ox$ (d) & 61.47 & 60.89 & 5.4  & 5.4    \\
\end{tabular}
\end{ruledtabular}
\caption{\label{tab:V} Original parameters of the real part of the
nucleon-nucleus optical potential \cite{Watson69} (first line)
and those adjusted to the energies of bound states or resonances in given
partial waves, all in units of MeV. The strength of the central part
is related to $v_R$ as 
$V_R = v_R  + 0.4ZA^{-1/3} \pm 27.0(N-Z)/A -0.3E_{\cm}\Theta(E_{\cm})$ 
and $V_{so}$ is the strength of the spin-orbit part;
see Ref.~\cite{Watson69} for more details.
}
\end{table}

Results for these studies are shown by the dotted  curves (M1) in
\Figs~\ref{fig:d12C-d12C}--\ref{fig:p13C-d12C} for $d + \C$ and $ p + \Cn$
 and \Figs~\ref{fig:d16O-d16O}--\ref{fig:p17O-d16O} for
$d + \Ox$ and  $p + \On$ at different energies. As
mentioned above the results shown by the dotted  curves in
\Fig~\ref{fig:d12C-d12C} (\Fig~\ref{fig:d16O-d16O}) are obtained with
a different Hamiltonian from those in
\Figs~\ref{fig:p13C-p13C}--\ref{fig:p13C-d12C}
(\Figs~\ref{fig:p17O-p17O}--\ref{fig:p17O-d16O}). In general the
description of the data for elastic scattering is fairly reasonable
and within  what can be expected from corresponding CDCC calculations.
For the transfer reactions $ p + \Cn \to d + \C$ 
and  $p + \On \to d + \Ox$ shown in
\Fig~\ref{fig:p13C-d12C} and \Fig~\ref{fig:p17O-d16O} respectively,
one gets a reasonable agreement with data in the forward direction
(except for a scaling factor), but deviations from data increase for
$\Theta_{\cm} > 30^{\circ}$. 
\renewcommand{\scl}{0.62}
\begin{figure}[!]
\begin{center}
\includegraphics[scale=\scl]{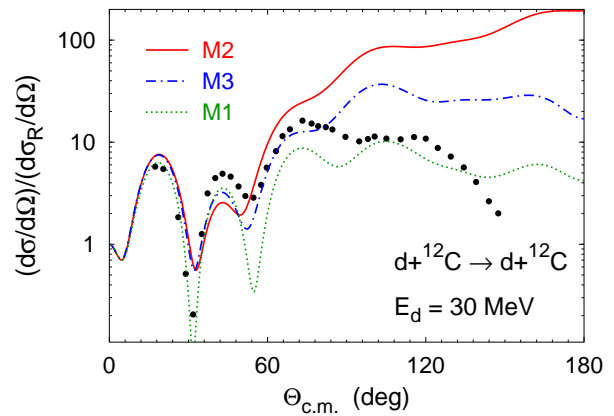}
\end{center}
\caption{\label{fig:d12C-d12C}  (Color online)
Differential cross section divided by Rutherford cross section
for  $d + \C$ elastic scattering at $E_d = 30$ MeV. 
Predictions of Model 1 (dotted curve),  Model 2 (solid curve), and  
Model 3 (dashed-dotted curve) are compared with 
the experimental data are from \Ref~\cite{dC30}.}
\end{figure}
\begin{figure}[!]
\begin{center}
\includegraphics[scale=\scl]{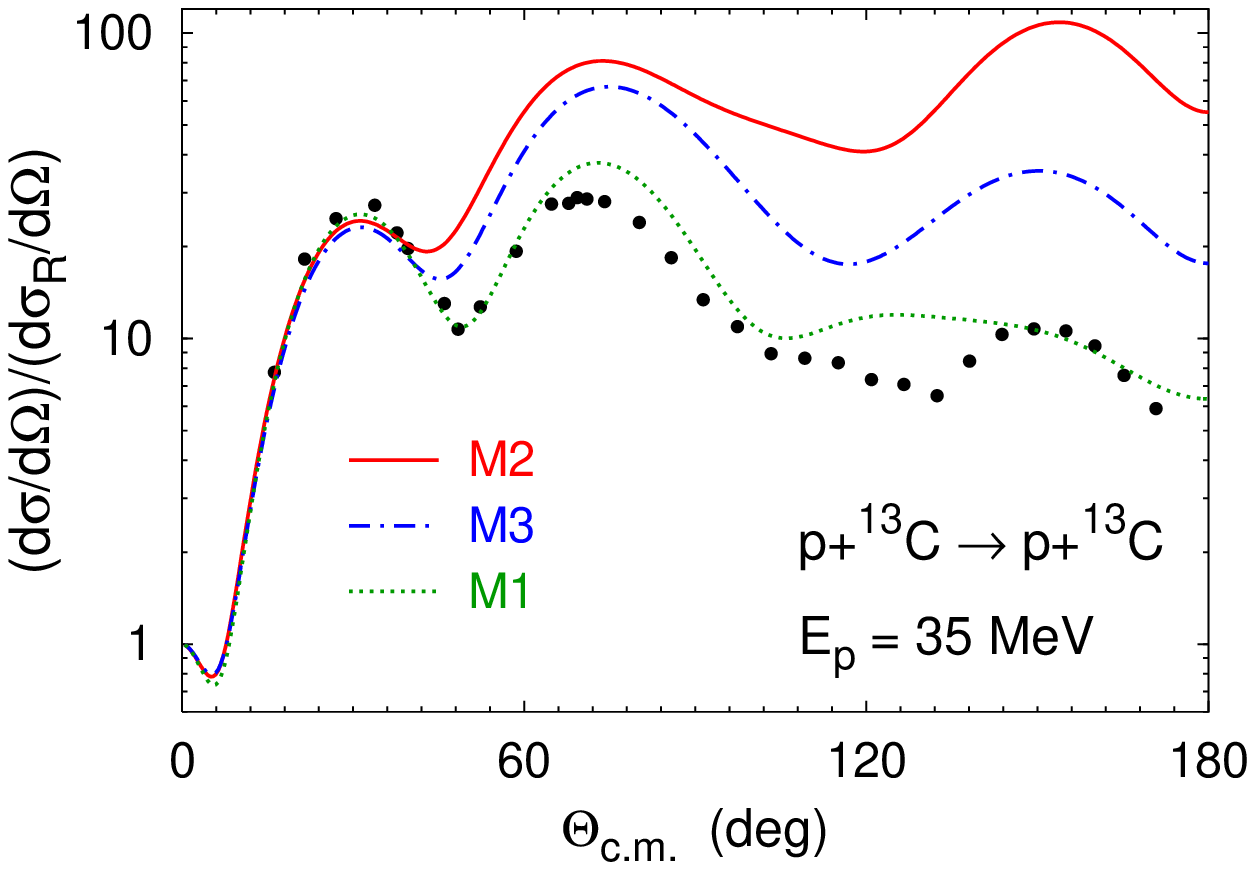}
\end{center}
\caption{\label{fig:p13C-p13C}  (Color online)
Differential cross section divided by Rutherford cross section
for  $p + \Cn$ elastic scattering at $E_p = 35$ MeV. 
Curves as in Fig.~\ref{fig:d12C-d12C}.
The experimental data are from \Ref~\cite{pCO35}.}
\end{figure}
\begin{figure}[!]
\begin{center}
\includegraphics[scale=\scl]{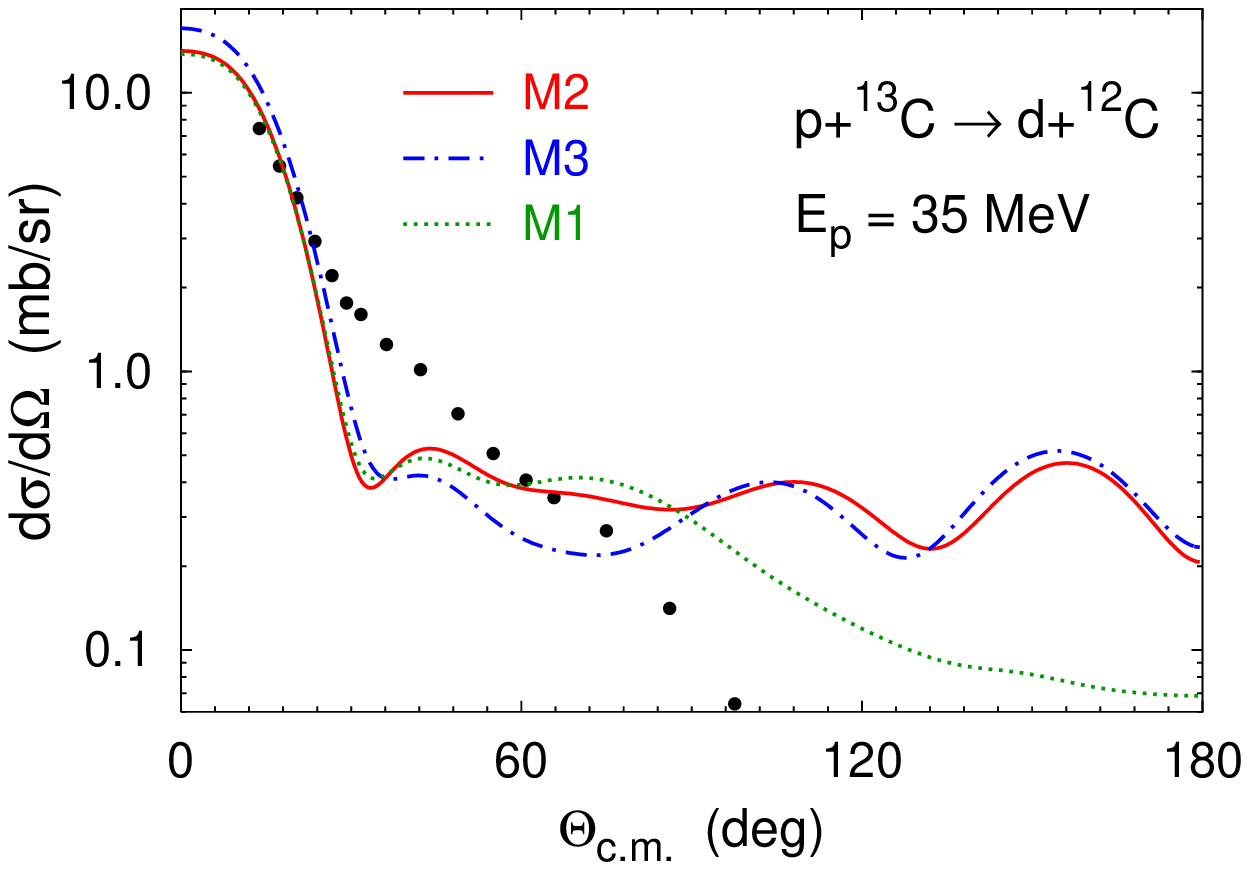}
\end{center}
\caption{\label{fig:p13C-d12C}  (Color online)
Differential cross section for  $p + \Cn \to d+\C$ transfer at $E_p = 35$ MeV. 
Curves as in Fig.~\ref{fig:d12C-d12C}.
The experimental data are from \Ref~\cite{pC35d}.}
\end{figure}
\begin{figure}[!]
\begin{center}
\includegraphics[scale=\scl]{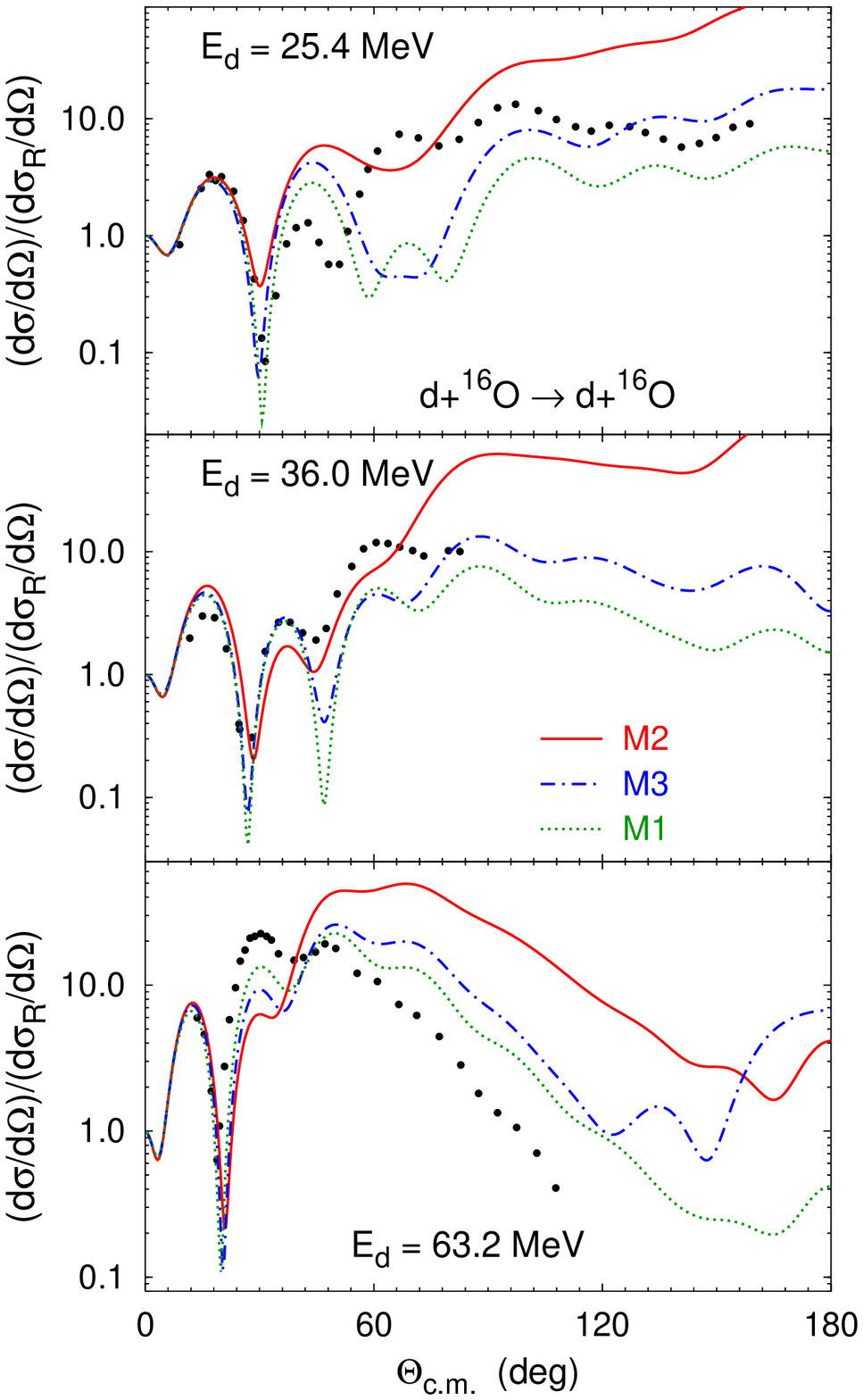}
\end{center}
\caption{\label{fig:d16O-d16O}  (Color online)
Differential cross section divided by Rutherford cross section
for  $d + \Ox$ elastic scattering at $E_d =$ 25.4, 36.0, and 63.2 MeV. 
Curves as in Fig.~\ref{fig:d12C-d12C}.
The experimental data are from \Refs~\cite{dO25-63,dO34}.}
\end{figure}
\begin{figure}[!]
\begin{center}
\includegraphics[scale=\scl]{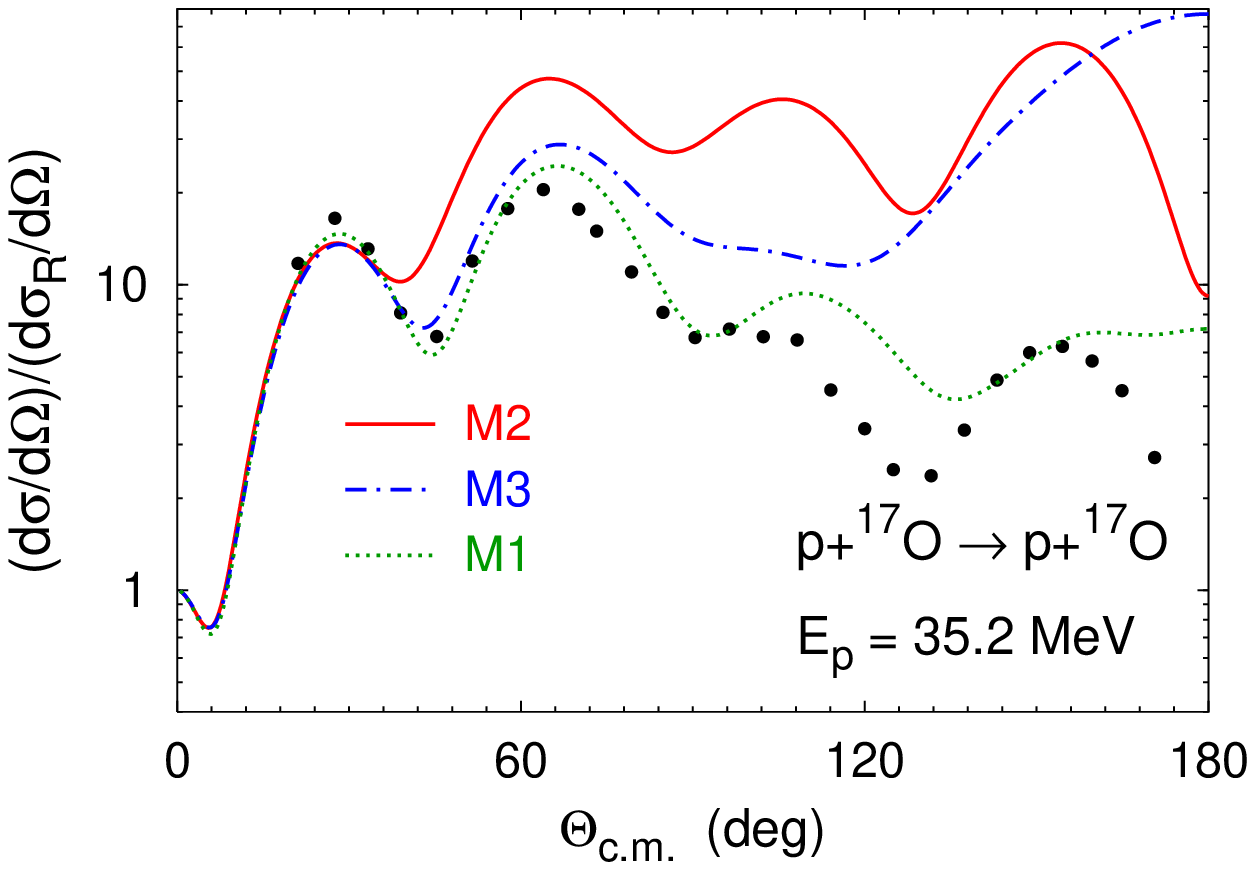}
\end{center}
\caption{\label{fig:p17O-p17O}  (Color online)
Differential cross section divided by Rutherford cross section
for  $p + \On$ elastic scattering at $E_p = 35.2$ MeV. 
Curves as in Fig.~\ref{fig:d12C-d12C}.
The experimental data are from \Ref~\cite{pCO35}.}
\end{figure}
\begin{figure}[!]
\begin{center}
\includegraphics[scale=\scl]{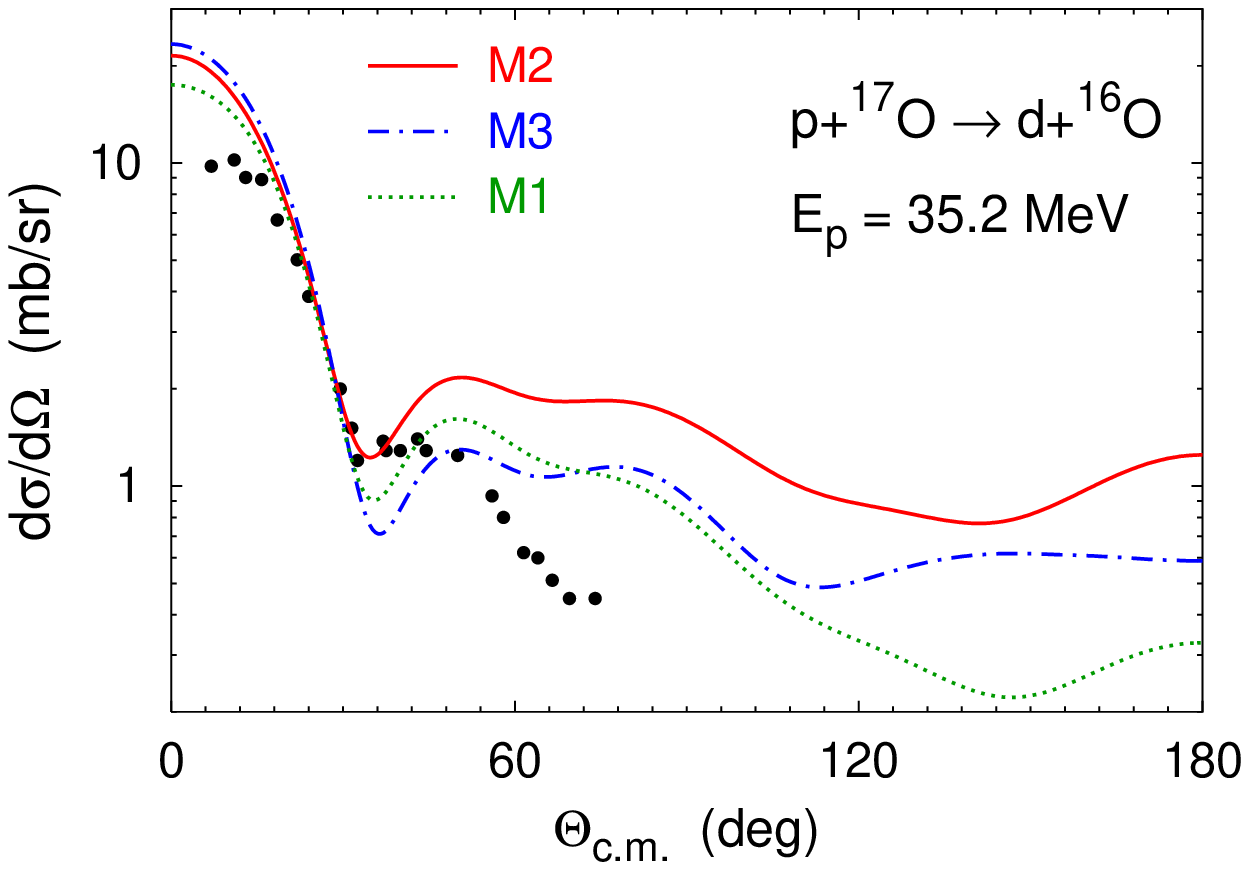}
\end{center}
\caption{\label{fig:p17O-d16O}  (Color online)
Differential cross section for $p+\On \to d+\Ox$ transfer at $E_p = 35.2$ MeV. 
Curves as in Fig.~\ref{fig:d12C-d12C}.
The experimental data are from \Ref~\cite{dO25-63}.}
\end{figure}

The fact that traditional three-body models of $d+A$ and $p + (An)$
scattering are inconsistent with each other encouraged us to study other
possibilities in order to shed light on the sensitivity of results to
different dynamical approaches.


\subsection{Model 2 - Energy-dependent optical potentials \label{sec:M2}}

The two-body t-matrix given by \Eq~\eqref{eq:t_alpha} enters the
 Faddeev/AGS equation \eqref{eq:Uba} for the transition
operator $U_{\beta \alpha}(Z)$. Even if the potential is energy
independent, the pair t-matrix has to be calculated
at the two-body energies $e = E - q^2_{\alpha}/2\mu_{\alpha}$, where
$q_{\alpha}$ is the relative momentum between particle $\alpha$ and
the $\cm$ of pair $\alpha$ that has to be integrated over
when solving the Faddeev/AGS equation, $\mu_{\alpha}$ is the corresponding
particle-pair $\alpha$ reduced mass, and
$E$ is three-body energy in the $\cm$ system.
Therefore in three-body calculations the particles in all pairs
scatter at two-body energies between $E$ and $-\infty$. In
the case of the CD-Bonn potential $np$ observables are described with
$\chi^2/\mathrm{datum} \sim 1$ from zero $np$ relative energy to the
$\pi$ production threshold. The same cannot be said about the $nA$ and
$pA$ optical potentials which in the previous model were chosen at a
fixed energy. Hence they describe the corresponding data at that
energy but not over the broader range that is relevant for the
solution of the three-body  Faddeev/AGS equation.

In the present model we take the full energy dependence of the optical
potential such that when $nA$ or $pA$ pairs interact at a given positive
relative energy, the used parameters of the optical potential
fit elastic $nA$ and $pA$ scattering at that energy. 
In addition, when the energy becomes negative the
corresponding potentials become real, energy-independent and support a
number of bound states that correspond to the ground and excited
states of the $(An)$ and $(Ap)$ nucleus whereas the Pauli forbidden
states are removed. As mentioned before, the parameters of the
energy-dependent optical potentials are slightly modified to obtain
the experimental binding energies at zero energy as indicated in 
Table~\ref{tab:V} for both  $nA$ or $pA$ potentials in given partial waves.
In addition, the binding energy of the Pauli forbidden $1p_{3/2}$ state 
in  $\Cn$ and $\A{13}{N}$ systems is fitted to the $\C$ neutron and proton 
separation energy, respectively, whereas the  $1p_{1/2}$
binding energy in $\On$ and $\A{17}{F}$ systems calculated with original
parameters \cite{Watson69} is close to the 
corresponding nucleon separation energies of $\Ox$. 
 The resulting  binding energies are given
in Table~\ref{tab:EB} for $\Cn$, $\A{13}{N}$, $\On$, and  $\A{17}{F}$ nuclei.
In the case of $N$-$\C$, where the adjusted parameters are quite 
different from the original ones, at positive energies  $v_R$ is replaced by
$v_R(E_{\cm}) = 60.0+(v_R-60.0)\exp(-E_{\cm}/2)$ and 
$V_{so}$ is replaced by $V_{so}(E_{\cm}) = 5.5+(V_{so}-5.5)\exp(-E_{\cm}/2)$,
such that the potential preserves the description of the $N$-$\C$
scattering data in the desired energy regime
and remains  a continuous function of the energy.
Such a replacement is not needed in the case of $N$-$\Ox$ where the 
adjusted parameters are very close to the original ones.

Using energy-dependent pair interactions in three-body calculations is
by no means free of theoretical complications, such as the problem of
non-orthogonality of three-body wave functions at different energies as
a result of the absence of a Hamiltonian theory for the scattering
process. This issue can be easily understood even at the two-body
level. If the potential is energy dependent the two-body bound states
and scattering states are not necessarily orthogonal, much like
scattering states corresponding to different energies. Therefore
completeness relations and three-particle unitarity may be at fault
even in the presence of real interactions. Nevertheless present
optical model fits, in particular the one by Watson~\etal, are rather
weak in their energy dependence, as can be seen by the strength of the
energy dependent coefficients vis-\`{a}-vis the energy independent
parameters; furthermore this energy dependence is smooth over the
energy range of the fit except perhaps the $N$-$\C$ spin-orbit interaction 
in  p waves near $e=0$. Even in this case we tried a different
p-wave interactions and the results are not very different as demonstrated
in the Appendix.  For this reason we believe that the problems
of non-orthogonality of wave functions and completeness may be
sufficiently small to allow a serious consideration of this model
given its notorious advantages
such as consistent dynamics for both $d+A$ and $p+(An)$ scattering
and the possibility to calculate transfer reactions to
$p+(An)$ and $n+(Ap)$ final states.

Furthermore, one should keep in mind that the energy dependence
and the imaginary part of the optical potential have the same origin;
they arise after the elimination of active degrees of freedom, i.e., 
excitations, multiconfiguration mixing,
 and breakup of nucleus $A$, from the considered Hilbert space
as described earlier by Feshbach \cite{feshbach}.
However, in a three-body system this leads in addition
to an effective energy-dependent  complex three-body potential,
 or, in general, to many-body potentials (up to $n$-body)
 in an $n$-body system as formally developed by Polyzou and Redish
\cite{polyzou:79a} in the framework of exact $n$-body theory. Well known
examples are the three- and four-nucleon systems described within the
interaction model with  energy-independent two-body potentials
allowing for an explicit excitation 
of a nucleon to a $\Delta$ isobar \cite{deltuva:03b,deltuva:08a}
which yields effective energy-dependent two-nucleon and many-nucleon forces
that are mutually consistent. In the study of three-nucleon observables
it was found that the $\Delta$-isobar effect of the two-nucleon
nature is often overcompensated by the  three-nucleon force effect.
Thus, also in the three-body nuclear reactions one could expect a similar
situation for some observables, that is, a partial cancellation of the effects 
arising due to the energy dependence of the two-body potential
and due to the three-body potential if the latter would be included in
the calculations. However, when the  energy dependence of the
two-body optical potential is introduced in the usual phenomenological way,
it is not clear at all what should be the consistent three-body potential.
We therefore do not attempt to include an optical three-body potential
in the present calculations, although such an extension of the 
Faddeev/AGS framework is possible.

\begin{table} [htbp]
\begin{ruledtabular}
\begin{tabular}{l*{5}{l}}
 & $1s_{1/2}$  & $2s_{1/2}$  & $1p_{3/2}$  & $1p_{1/2}$  & $1d_{5/2}$ \\ \hline
$\A{13}{C}$ & 38.022* & 1.857 & 18.722* & 4.946 & 1.092 \\
$\A{13}{N}$ & 33.864* &  & 15.957* & 1.944 &  \\
$\A{17}{O}$ & 37.213* & 3.272 & 19.267* & 16.067* & 4.143 \\
$\A{17}{F}$ & 32.559* & 0.105 & 15.561* & 12.348* & 0.600 \\
\end{tabular}
\end{ruledtabular}
\caption{\label{tab:EB} Binding energies (MeV) of the bound states corresponding
to the potential parameters of Table~\ref{tab:V}.
Pauli forbidden bound states that are removed are marked with *.
}
\end{table}

\renewcommand{\scl}{0.62}
\begin{figure}[!]
\begin{center}
\includegraphics[scale=\scl]{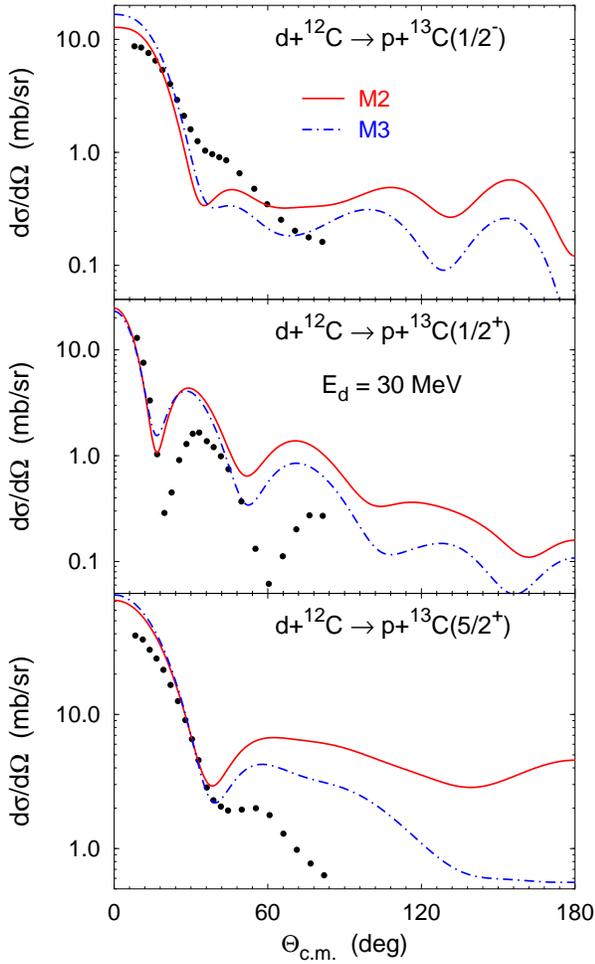}
\end{center}
\caption{\label{fig:d12C-p13C}  (Color online)
Differential cross section for  $d + \C \to p+\Cn$ transfer at $E_d = 30$ MeV. 
Predictions of Model 2 (solid curve) and  
Model 3 (dashed-dotted curve) are compared with 
the experimental data from \Ref~\cite{dC30p}.}
\end{figure}
\begin{figure}[!]
\begin{center}
\includegraphics[scale=\scl]{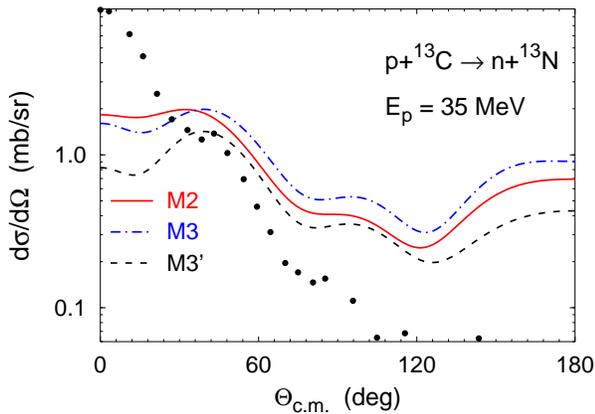}
\end{center}
\caption{\label{fig:p13C-n13N}  (Color online)
Differential cross section for  $p + \Cn \to n+ \A{13}{N}$ reaction at 
$E_p = 35$ MeV. Dashed curve is the prediction of Model 3', other
curves as in Fig.~\ref{fig:d12C-p13C}.
The experimental data are from \Ref~\cite{pC35n}.}
\end{figure}
\begin{figure}[!]
\begin{center}
\includegraphics[scale=\scl]{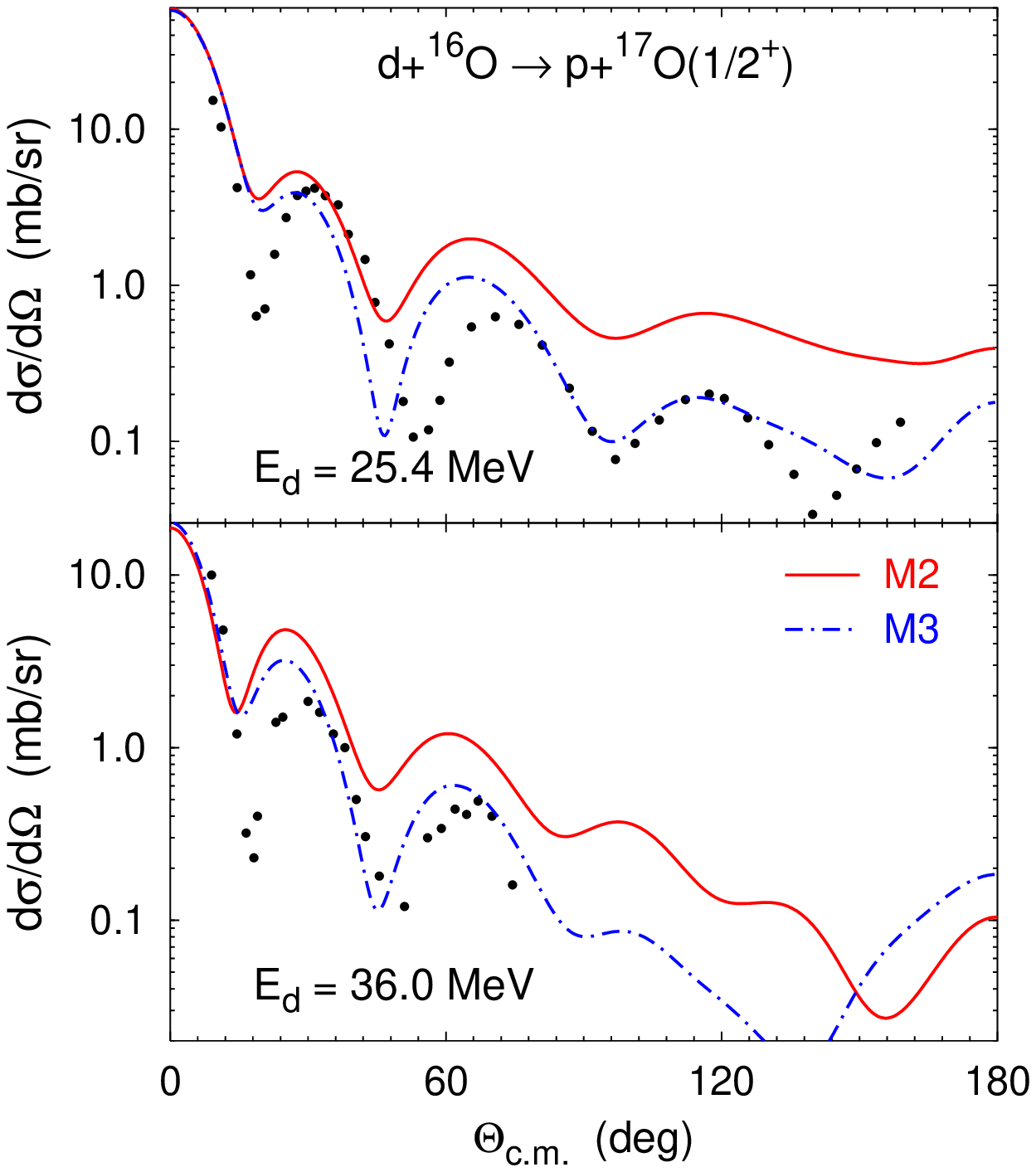}
\end{center}
\caption{\label{fig:d16O-p17Os}  (Color online)
Differential cross section for  $d + \Ox \to p+\On$ transfer at 
$E_d =$ 25.4 and 36.0 MeV. 
Curves as in Fig.~\ref{fig:d12C-p13C}.
The experimental data are from \Ref~\cite{dO25-63}.}
\end{figure}
\begin{figure}[!]
\begin{center}
\includegraphics[scale=\scl]{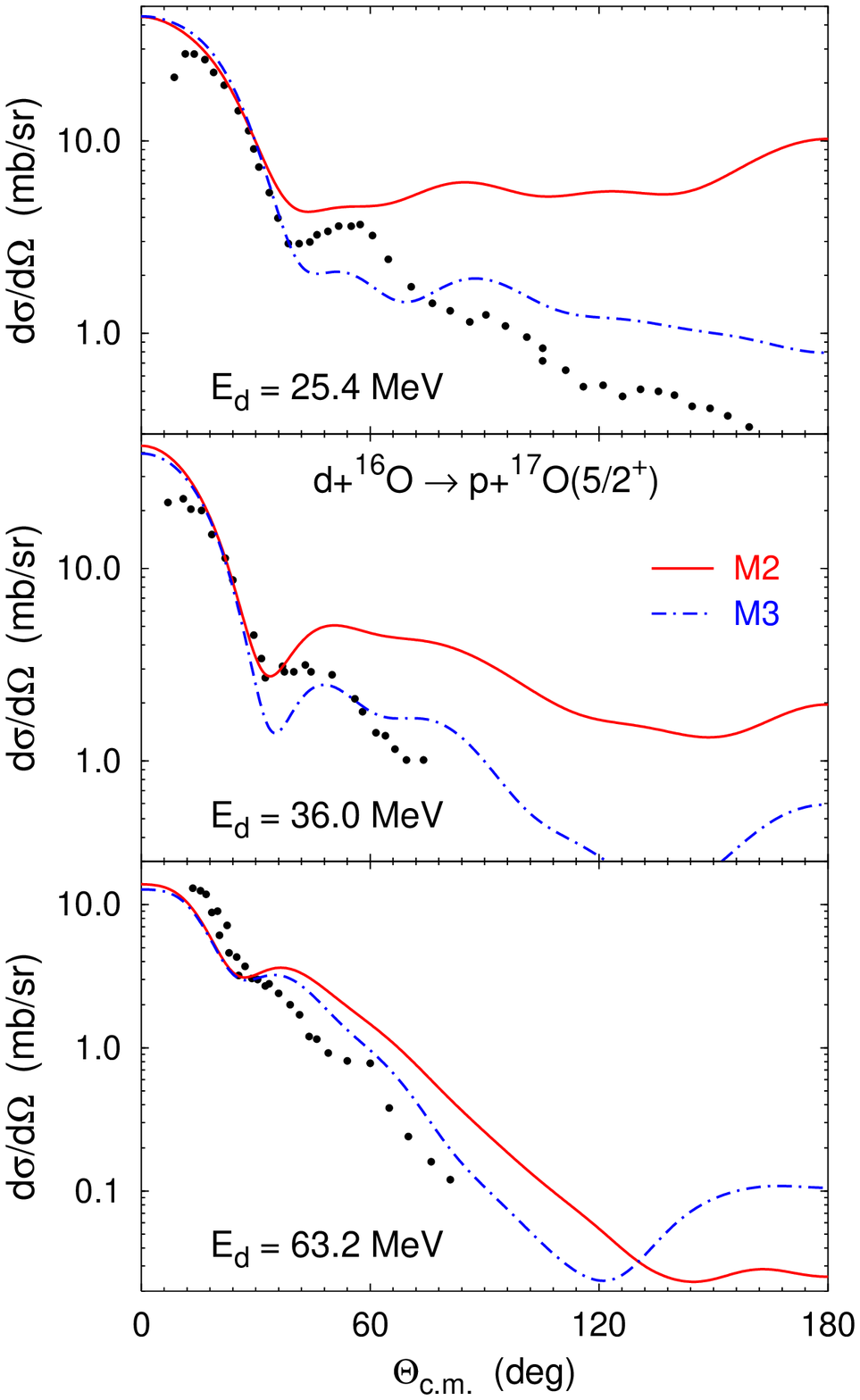}
\end{center}
\caption{\label{fig:d16O-p17Od}  (Color online)
Differential cross section for  $d + \Ox \to p+\On$ transfer at 
$E_d =$ 25.4, 36.0, and 63.2 MeV. 
Curves as in Fig.~\ref{fig:d12C-p13C}.
The experimental data are from \Ref~\cite{dO25-63}.} 
\end{figure}

\begin{figure}[!]
\begin{center}
\includegraphics[scale=\scl]{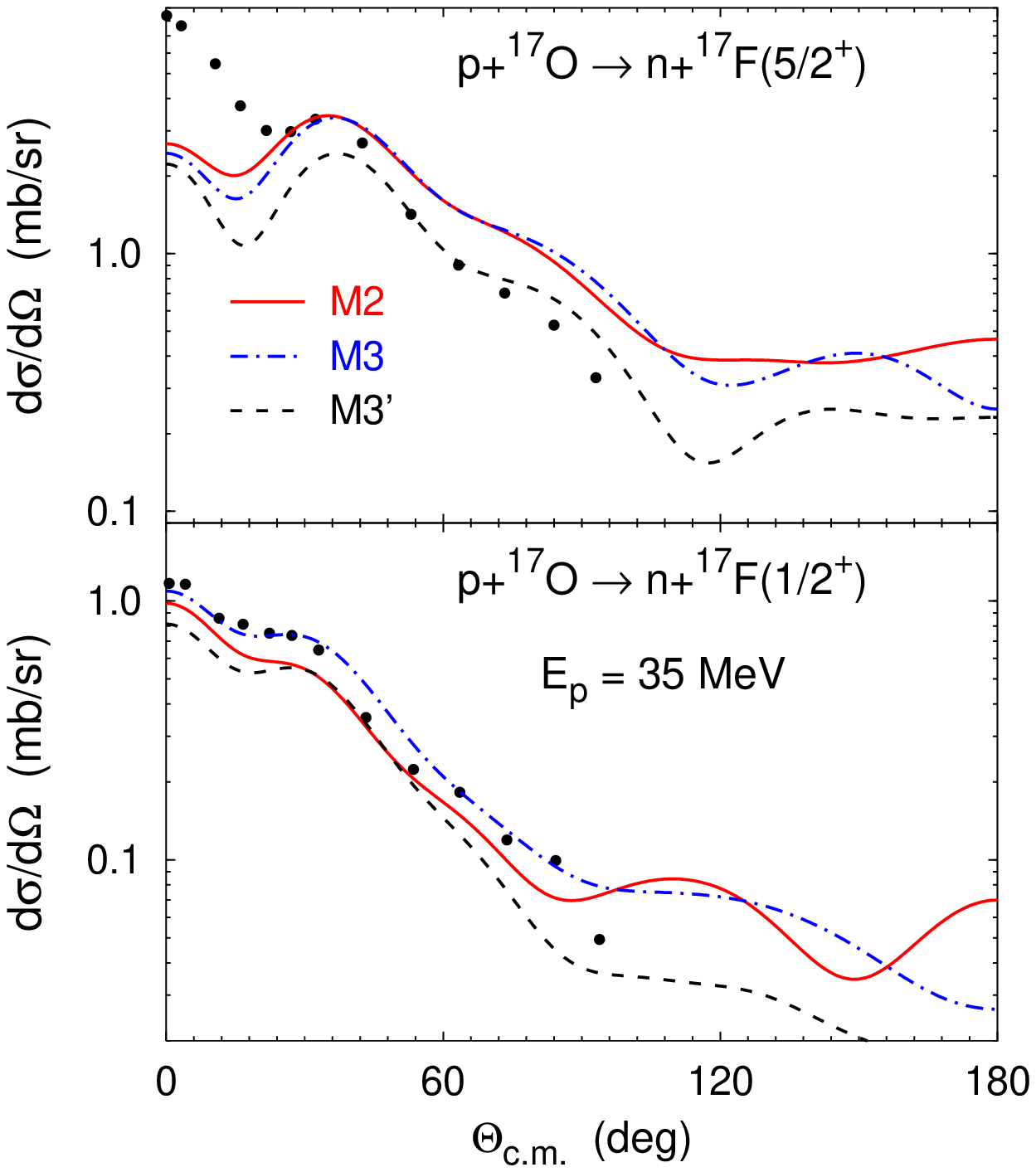}
\end{center}
\caption{\label{fig:p17O-n17F}  (Color online)
Differential cross section for  $p + \On \to n+ \A{17}{F}$ reaction at 
$E_p = 35$ MeV. Curves as in Fig.~\ref{fig:p13C-n13N}.
The experimental data are from \Ref~\cite{pO35n}.}
\end{figure}

In \Fig~\ref{fig:d12C-d12C}--\ref{fig:p17O-n17F} the solid curves (M2)
show the results of the present fully energy-dependent model for all
possible reactions at different energies. A number of interesting
features emerge: 

\begin{itemize}
\item[a)]  Elastic scattering results shown in
  \Figs~\ref{fig:d12C-d12C}, \ref{fig:p13C-p13C}, \ref{fig:d16O-d16O},
and \ref{fig:p17O-p17O} differ quite strongly
  from Model 1 (dotted curves), particularly at large angles, and
 become considerably worse when compared  to data.

\item[b)] In the low angular region $(\Theta_{\cm} < 30^{\circ})$ 
$p + \Cn \to d + \C$ (\Fig~\ref{fig:p13C-d12C}) and 
$p + \On \to d + \Ox$ (\Fig~\ref{fig:p17O-d16O})
  results are very similar to those obtained with Model 1 except for a
  small scaling factor. 

\item [c)] \Fig~\ref{fig:d12C-p13C} shows new results for the transfer
  reactions $d + \C \to p + \Cn$ to  ground state $1/2^-$ and
   excited states $1/2^+$ and $5/2^+$. Again up to $\Theta_{\cm} \simeq
  30^{\circ}$ the calculation follows the data within a small scaling
  coefficient that may be associated with a spectroscopic factor. In
  the case of the transfer to the ground state, solid curves in
  \Figs~\ref{fig:p13C-d12C} and \ref{fig:d12C-p13C}  
  have similar shape as expected by detailed balance taking into 
  account the small difference in the energies.
 The calculations also reflect the qualitative  features of the data. 

\item[d)] \Figs~\ref{fig:d16O-p17Os} and \ref{fig:d16O-p17Od} show new
  results for the transfer reactions $d + \Ox \to p + \On$ to
   ground state $ 5/2^+$  and  excited state $1/2^+$. Again
  the calculations describe the qualitative features of the data
  though scaling factors may be needed.

\item[e)] \Figs~\ref{fig:p13C-n13N} and \ref{fig:p17O-n17F} show new
  results for $p + \Cn  \to n + \A{13}{N}$   ground state $ 1/2^-$
  and $p + \On  \to n + \A{17}{F}$   ground  state $5/2^+$
and  excited state $1/2^+$. Although in the charge exchange
  reactions to the ground state the data is not described successfully,
  it is worth noting that in $p + \On \to n + \A{17}{F}$ excited state 
  $(1/2^+)$ the calculations are in very reasonable agreement with
  data, except for a small scaling factor. 

\end{itemize}

It is worth noting at this point that a good description of elastic data 
beyond small angles does not seem to be necessary to get the right magnitude 
of the transfer cross sections at small angles since
\Figs~\ref{fig:p13C-d12C} and \ref{fig:p17O-d16O} show similar results
for two distinct models that lead to very different results for the
elastic cross sections at large angles
(see \Figs~\ref{fig:d12C-d12C}, \ref{fig:p13C-p13C}, 
\ref{fig:d16O-d16O}, \ref{fig:p17O-p17O}).   


\subsection{Model 3 - A ``hybrid" optical potential approach}

Having studied these two extreme dynamical model approaches, the
energy-independent and the fully energy-dependent, we attempt to study
a combination of the two. Since we want that the relevant nuclei,
 $\Cn$, $\A{13}{N}$, $\On$, and  $\A{17}{F}$
have the proper low-energy spectra  in order to describe
all the relevant transfer reactions discussed before, we use in this case a
partial-wave dependent optical potential in the following way: 
{\em a)} For $d +A$ reactions in $N$-$\C$  ($N$-$\Ox$) s, p, and d waves 
 (s and d waves)  we use
the energy-dependent optical potentials of Model 2; 
for $p+(An)$ reactions the $pA$ potential in the above mentioned partial 
waves is energy-dependent as well, but the $nA$ potential is taken over
from Model 1 since it is sufficient to bind $\Cn$ and $\On$;
{\em b)} In all other partial waves we use the energy-independent optical
potentials of Model 1 with a few nuances that are explained in the
text, depending on whether we have $d+A$ or $p+(An)$ scattering.

Since Model 1 is more absorptive than Model 2 due to the large impact
of the imaginary part of the optical interactions on the elastic cross
sections we expect this hybrid model to improve the description of the
elastic data. 

For $d+A$ scattering, results are shown by the dash-dotted
curves (M3) in \Figs~\ref{fig:d12C-d12C}, \ref{fig:d16O-d16O}, 
\ref{fig:d12C-p13C}, \ref{fig:d16O-p17Os}, and
and \ref{fig:d16O-p17Od}. In $d+\C \; (d+\Ox)$ both $nA$ and
$pA$ optical potentials are, like in Model 2, energy dependent
in s, p, and d waves (s and d waves) while in all other partial
waves they are energy independent with the parameters chosen at half the
deuteron lab energy like in Model 1. The dash-dotted curves show a
remarkable improvement  vis-\`{a}-vis the fully energy-dependent
calculations (solid  lines in Model 2), particularly at large
angles. This effect is visible not only in elastic scattering
(\Figs~\ref{fig:d12C-d12C} and \ref{fig:d16O-d16O}), but also 
in the transfer reactions $d+A \to p+(An)$ shown in
\Figs~\ref{fig:d12C-p13C}, \ref{fig:d16O-p17Os}, and \ref{fig:d16O-p17Od}
where in some specific cases such as in
\Figs~\ref{fig:d16O-p17Os} and \ref{fig:d16O-p17Od} one gets quite
reasonable description of the data.

For $p+(An)$ scattering results are again shown in
\Figs~\ref{fig:p13C-p13C}, \ref{fig:p13C-d12C}, \ref{fig:p17O-p17O}, 
\ref{fig:p17O-d16O}, \ref{fig:p13C-n13N},
and \ref{fig:p17O-n17F} by the dash-dotted curves (M3). In
$p+\Cn \; (p+\On)$ the $pA$ optical potentials are,
like in Model 2, energy dependent in s, p, and d waves (s and
d waves) and, in all other partial waves, are energy independent with
the parameters chosen at the proton lab energy, like in Model 1. As
for the $nA$ optical potential it is chosen as in Model 1 where,
in all partial waves the potential is real and supports a number of
single particle states as mentioned before. As in $d+A$ reactions, we
notice an improvement in the description of $p+(An)$ elastic 
(\Figs~\ref{fig:p13C-p13C} and \ref{fig:p17O-p17O}) as well as $p+(An)
\to d+A$ transfer (\Figs~\ref{fig:p13C-d12C}  and \ref{fig:p17O-d16O})
observables. Nevertheless, at
small angles $(\Theta_{\cm} \leq 30^{\circ})$, the differences between
Models 1, 2, and 3 are quite small indicating that the extracted 
spectroscopic factors would be of similar size as well.

Finally for the charge transfer reactions shown in
\Figs~\ref{fig:p13C-n13N} and \ref{fig:p17O-n17F} we add a new calculation
shown by the dashed  curves (M3') where the $nA$ optical
potential is energy-dependent in the s, p, and d partial waves for
$\Cn$ and s and d waves for $\On$ like in Model 2 but
is, in the other partial waves, energy-independent with the parameters
chosen according to the lab energy of the neutron in the 
inverse reaction $n + (Ap) \to p + (An)$. The three
curves shown in \Figs~\ref{fig:p13C-n13N} and \ref{fig:p17O-n17F} are not
very different aside a scaling factor.


\section{Distorted Wave Equations}\label{sec:DWE}

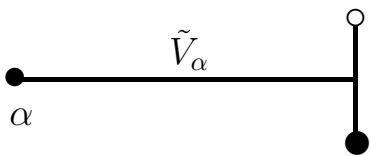
\begin{figure} [b]
\begin{picture}(150,40)
\linethickness{0.5mm} \put(7.0,6.3){\makebox(246,33){\LARGE{$\circ$}}}
\put(7.0,6.3){\makebox(120,10){{\Large $\tilde V_{\alpha}$}}}
\put(1,1){\circle*{7.0}}\put(0,0){\line(1,0){130}}
\put(130,0){\line(0,1){20}} \put(130,0){\line(0,-1){20}}
\put(1,1){\makebox(5,-30){{\Large $\alpha$}}}
\put(7.0,6.4){\makebox(256,-61){\circle*{9.0}}}
\end{picture}
\vspace{0.7cm}
\caption{$\tilde V_{\alpha}$ optical interaction in channel $\alpha$} 
\label{fig:Vcm}
\end{figure}

In order to relate our calculations to the standard approaches of
nuclear reaction theory we derive an alternative set of scattering equations.
Introducing an effective interaction $\tilde V_{\alpha}$ acting
between particle $\alpha$ and the $\cm$ of pair $\alpha$ as shown in
\Fig~\ref{fig:Vcm} one may define a new resolvent  
\begin{gather} \label{eq:G_alpha}
\tilde G_{\alpha}(Z) = (Z - H_0 - v_{\alpha} - \tilde V_{\alpha})^{-1},
\end{gather}
such that
\begin{gather}\label{eq:tildeGZ}
\begin{align}
\tilde G_{\alpha}(Z) =  {} & G_{\alpha}(Z) + G_{\alpha}(Z) \; \tilde
T_{\alpha} (Z) \; G_{\alpha}(Z), \\
\tilde T_{\alpha} (Z) =  {} & \tilde V_{\alpha}  + \tilde V_{\alpha} \;
G_{\alpha}(Z) \; \tilde T_{\alpha} (Z).
\end{align}
\end{gather}
Likewise one may define a distorted wave in channel $\alpha$ as 
\begin{gather}
|\tilde \psi_{\alpha} \rangle = (1 + G_{\alpha} (Z) \; \tilde T_{\alpha}(Z))
| \psi_{\alpha} \rangle.
\end{gather}
Using the identity
\begin{gather}
G(Z) =  \tilde G_{\beta}(Z) + \tilde G_{\beta}(Z) 
 [ \tilde G_{\beta}^{-1}(Z)  -  G^{-1}(Z)  ] G(Z),
\end{gather}
together with \Eqs~\eqref{eq:G} and \eqref{eq:G_alpha} one gets 
\begin{gather}
G(Z) =  \tilde G_{\beta}(Z) + \tilde G_{\beta}(Z) \, \Omega_{\beta} \, G(Z),
\end{gather}
where
\begin{gather}\label{eq:Vb}
\Omega_{\beta} = \sum_{\sigma}  \bar{\delta}_{\sigma \beta} \; v_{\sigma} -  
\tilde V_{\beta}. 
\end{gather}
A  new operator $\tilde U_{\beta \alpha}(Z)$ 
relating $G(Z)$ to $\tilde G_{\alpha} (Z)$ instead of $G_{\alpha}(Z)$, i.e.,
\begin{gather} \label{eq:G=delta}
G(Z) =  \delta_{\beta \alpha}\,  \tilde G_{\alpha} (Z) +  \tilde
G_{\beta} (Z) \, \tilde U_{\beta \alpha} (Z)  \, \tilde G_{\alpha} (Z),
\end{gather}
satisfies an equation 
\begin{gather} \label{eq:tildeUba}
\tilde U_{\beta \alpha} (Z) =  \bar{\delta}_{\beta \alpha} \tilde
G_{\alpha}^{-1} (Z) + \Omega_{\beta} + \Omega_{\beta} \, \tilde
G_{\alpha} (Z) \, \tilde U_{\alpha \alpha} (Z).
\end{gather}
Its relation  to the standard Faddeev/AGS operator $U_{\beta \alpha} (Z)$ is 
obtained using \Eq~\eqref{eq:tildeGZ} in \eqref{eq:G=delta} and comparing back 
with \eqref{eq:G(Z)} as
\begin{gather} \label{eq:UDelta}
\begin{split}
U_{\beta \alpha} (Z) = {} & \delta_{\beta \alpha}\; \tilde T_{\alpha}(Z) + 
\left[1 + \tilde T_{\beta} (Z) \; G_{\beta} (Z)\right]  \\   & \times
\tilde U_{\beta \alpha} (Z)  \left[1 +  G_{\alpha} (Z)\;
  \tilde T_{\alpha} (Z)\right],
\end{split}
\end{gather}
which for on-shell elements reads
\begin{gather} \label{eq:Ubetalph}
\langle\psi_{\beta}|U_{\beta \alpha} (Z) |\psi_{\alpha}\rangle = 
 \delta_{\beta \alpha} 
\langle\psi_{\beta}|\tilde{T}_{\alpha} (Z) |\psi_{\alpha}\rangle +
\langle \tilde \psi_{\beta} |  \tilde U_{\beta \alpha} (Z) 
| \tilde \psi_{\alpha}\rangle.
\end{gather}
In \Eq~\eqref{eq:tildeUba} the term 
$\bar{\delta}_{\beta \alpha} \tilde G_{\alpha}^{-1}(Z)$ is zero on-shell
and will be omitted in the following considerations.

Using the Born approximation 
$\tilde U_{\beta \alpha}(Z) \simeq \Omega_{\beta}$ for  $\beta \neq \alpha$,
one gets
\begin{gather} \label{eq:Usimeq}
\langle\psi_{\beta}|U_{\beta \alpha} (Z) |\psi_{\alpha}\rangle \simeq \langle \tilde
\psi_{\beta} |  \Omega_{\beta } | \tilde \psi_{\alpha}\rangle,
\end{gather}
which corresponds to the usual distorted-wave Born approximation (DWBA) 
for the transfer reactions in the post form. 

On the other hand, using $\tilde V_1 = 0$ in the case of  particle 1 
colliding with pair (23) one gets $\tilde T_{\alpha}(Z)=0$,
$U_{11}(Z) = \tilde U_{11}(Z)$, and  $\Omega_1 = v_2 + v_3$ leading to
\begin{gather}
U_{11} (Z) = (v_2 + v_3) + (v_2 + v_3)\, G_1 (Z) \, U_{11}(Z), 
\label{eq:u11} 
\end{gather}
which is the integral form of the CDCC differential
equation. This equation by itself is not connected in all orders of iteration 
and therefore cannot be solved by standard numerical methods since it does 
not satisfy the Fredholm alternative. Nevertheless one may follow the 
momentum space version of the CDCC approach and  use the spectral 
decomposition of $G_1(Z)$ to obtain a set of coupled equations involving 
the continuum wave functions of pair (23) in addition to the bound state 
wave function $| \psi_{1}\rangle$.
If the continuum is discretized and the corresponding wave functions 
normalized \`a la CDCC, the solution of  \Eq~\eqref{eq:u11} includes the 
bound to continuum and continuum to continuum couplings 
that are common to CDCC calculations. 
In \Ref~\cite{deltuva:07d} we have shown that CDCC calculations for deuteron 
elastic scattering and breakup from a heavier target are reliable, but 
transfer and breakup reactions involving the scattering of a halo nucleus 
from a light target, such as  $\Bee + p$, may be at fault. 
Therefore we expect all deuteron elastic scattering results shown in 
\Figs~\ref{fig:d12C-d12C} and \ref{fig:d16O-d16O} to agree well with those 
obtained from equivalent CDCC calculations.

The present derivations may be useful in future studies of approximate 
methods often used in nuclear reaction calculations involving deuterons  
or halo nuclei. 
Since at this time in the present framework we do not have the means to test 
the validity of \Eq~\eqref{eq:Usimeq} or any other approximation,
we compare our results with published calculations involving either DWBA, 
coupled-channel Born approximation (CCBA), or various adiabatic approaches
\cite{Johnson70,Johnson97,Timo99}.
These calculations use wave functions $|\tilde{\psi}_\alpha \rangle$
and optical potentials $\tilde{V}_\alpha$ that are tuned at the 
considered reaction energies, while our calculations 
use global fits to nuclear reaction data and are aimed at providing a 
description of the data in different channels simultaneously.
That tuning may be, at least in part, the reason for a better description 
of the data as discussed below.

The data points in \Fig~\ref{fig:p13C-d12C} for $p+\Cn \to d+\C$ were 
analyzed in \Ref~\cite{pC35d} using DWBA, adiabatic deuteron breakup
approximation (ADBA), and CDCC-CCBA. 
Up to $30^{\circ}$ our results coincide with all of the 
previous calculations, but at larger angles adiabatic and CDCC-CCBA 
calculations follow the data much better than ours.

Likewise, the data points in \Fig~\ref{fig:d12C-p13C} for 
$d+\C \to p+\Cn$ were analyzed in \Ref~\cite{dC30p} using DWBA
for transition to the $\Cn$ ground state and CCBA 
for  transitions to the $1/2^+$ and $5/2^+$ excited states. 
While DWBA provides a better fit to the data at small angles, 
it still overshoots the data at larger angles much like our calculations. 
As for the reactions leading to the $1/2^+$ and $5/2^+$ excited states of 
$\Cn$, both CCBA and our calculations describe the data equally badly. 

In \Ref~\cite{pC35n}  the $p+\Cn \to n+\A{13}{N}$ reaction was analyzed 
with DWBA which undershoots the data at small angles, much like our results 
shown in \Fig~\ref{fig:p13C-n13N}, but, overall, 
provide a better description of the data.

The results in \Fig~\ref{fig:d16O-p17Os} and  \Fig~\ref{fig:d16O-p17Od} 
for $d+\Ox \to p+\On$ may be compared with DWBA studies from 
\Ref~\cite{dO25-63}  and adiabatic calculations from \Ref~\cite{Timo99} 
for 36 MeV and 63.2 MeV deuterons. The results we get in the framework 
of Model 3 (the ``hybrid") are qualitatively similar to those obtained 
in \Ref~\cite{dO25-63,Timo99} although quantitatively they may differ 
in specific angular regions, leading to a 
description of the data that is not as good as the one provided by DWBA 
or adiabatic calculations. 

Finally the data in \Fig~\ref{fig:p17O-n17F}  for 
$p+\On \to n+\A{17}{F}$ is analyzed in \Ref~\cite{pO35n} using DWBA. 
The DWBA calculations leading to the ground state of $\A{17}{F}$ are 
qualitatively similar to ours but fit the data  better at the 
forward angles. For the transition to the $1/2^+$ excited state of 
$\A{17}{F}$ both calculations are quantitatively similar and fit the 
data equally well.

\begin{figure}[!]
\begin{center}
\includegraphics[scale=0.54]{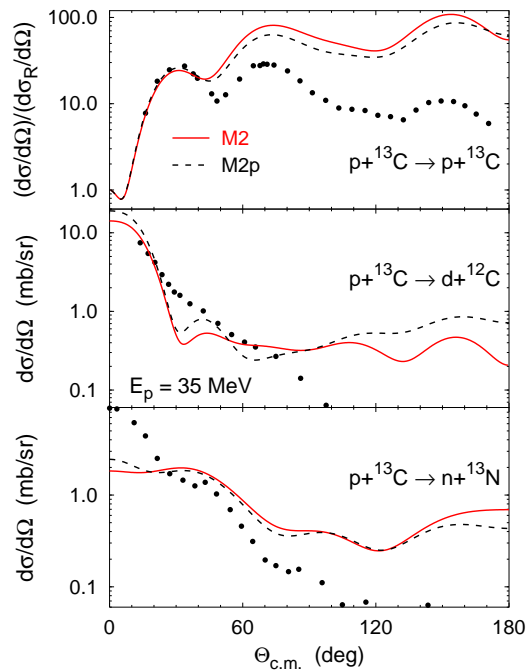}
\end{center}
\caption{\label{fig:M2p}  (Color online)
Differential cross section for  $p + \Cn$ elastic and transfer reactions
at $E_p = 35$ MeV. 
Predictions of Model 2 with  p-wave potential from  Table~\ref{tab:V}
(solid curves) and  \ref{tab:Cp} (dashed curves) are compared.
The experimental data are from \Refs~\cite{pCO35,pC35d,pC35n}.}
\end{figure}

\section{Conclusions}\label{sec:CON}

We have used the Faddeev/AGS three-body approach to study $d+\C$, $d+\Ox$, 
$p+\Cn$, and $d+\On$ reactions as a three-body system made up by a proton 
$p$, a neutron $n$ and a structureless nuclear core A, the $\C$ or the 
$\Ox$. The interactions between pairs are the realistic interactions that 
describe $np$, $nA$, and $pA$ scattering over the relevant energy range. 
The Coulomb interaction between the proton and the nuclear core is included 
in a numerically exact (converged) way. 

The aim of the present work is to demonstrate the possibilities and the 
shortcomings of the Faddeev/AGS
 three-body approach that provides, simultaneously, 
predictions for all possible reactions, i.e., elastic, transfer and charge 
exchange such as, for example, $p+\On \to p+\On$, $p+\On \to d+\Ox$, and 
$p+\On \to n+\A{17}{F}$, or $d+\Ox \to d+\Ox$,  $d+\Ox \to p+\On$,
and $d+\Ox \to n+\A{17}{F}$.

Three different models (M1, M2, M3) are studied involving 
energy-independent and energy-dependent optical potentials that fit the 
$nA$ and $pA$ elastic scattering and whose parameters are fixed at a 
chosen energy or are allowed to vary over the energy range of the 
interacting pair,  respectively. In the case of energy-dependent optical 
potentials these become real at negative energies and support a number of 
single particle states that characterize the $(An)$ or the $(Ap)$ nucleus.

The results of our calculations indicate that transfer and charge exchange
 reactions at small angles are rather insensitive to the chosen model, but 
 the elastic scattering cross sections are highly sensitive to the choice 
of energy dependence of the optical interaction (M1 versus M2 and M3). 
Comparison with published CDCC, DWBA, CCBA, and adiabatic calculations 
indicates that these approximate methods provide, in general, a better 
fit of the data than our calculations but are qualitatively similar to 
our results, particularly the ones of the ``hybrid" model M3 that uses a 
partial-wave dependent optical potential whose parameters are 
energy-independent except in the partial waves that support the 
single particle states of the $(An)$ and $(Ap)$ nuclei.


\begin{acknowledgments}
The authors thank F.~M.~Nunes and I.~J.~Thompson for the comments on the 
manuscript. A.D. is supported by the Funda\c{c}\~{a}o para a Ci\^{e}ncia 
e a Tecnologia (FCT) grant SFRH/BPD/34628/2007
and A.C.F. in part by the FCT grant POCTI/ISFL/2/275.
\end{acknowledgments}


\begin{appendix}
\section{}
We present here selected results obtained with an alternative $N$-$\C$ 
p-wave potential whose spin-orbit strength 
is the same as in the other partial waves (see Table~\ref{tab:V}).
The strength of the central part is adjusted to reproduce 
$\Cn$ and $\A{13}{N}$  $1p_{1/2}$ 
ground state energies as in Table~\ref{tab:EB},
thereby resulting different binding energies for the Pauli forbidden 
$1p_{3/2}$ bound states that are given in Table~\ref{tab:Cp}
together with the new values of potential parameters.
\begin{table} [htbp]
\begin{ruledtabular}
\begin{tabular}{*{5}{c}}
$v_R(nA)$ & $v_R(pA)$  & $V_{so}(NA)$ & 
$1p_{3/2}(\Cn)$ &  $1p_{3/2}(\A{13}{N})$  \\ \hline
49.61 & 49.11 & 5.5 & 8.587* & 5.507*
\end{tabular}
\end{ruledtabular}
\caption{\label{tab:Cp} 
Parameters of the alternative $N$-$\C$ p-wave potential
together with the resulting  binding energies for the Pauli forbidden 
$1p_{3/2}$ bound state. 
See Tables \ref{tab:V} and \ref{tab:EB} for further explanations.
}
\end{table}
The predictions of Model 2 with p-wave potential from Table~\ref{tab:V}
and  \ref{tab:Cp} are compared in Fig.~\ref{fig:M2p}
for $p + \Cn$ elastic and transfer reactions. The differences are
rather insignificant when compared to the discrepancies between theory and data
and therefore do not change the conclusions of this paper.
 Differences of similar magnitude can be seen also for the observables 
of $d + \C$ reactions.

\end{appendix}


\end{document}